\documentclass[final,5p,times,twocolumn]{elsarticle}

\usepackage{amssymb}
\usepackage[utf8]{inputenc} 
\usepackage[T1]{fontenc}
\usepackage{siunitx} 
\usepackage{xparse} 
\usepackage{physics} 
\usepackage{graphicx}
\usepackage{natbib,hyperref}

\journal{Journal of Magnetism and Magnetic Materials}

\begin{document}

\begin{frontmatter}



\title{Shaping magnetization dynamics in a planar square dot by adjusting its surface anisotropy}


\author[inst1]{Grzegorz Centa{\l}a}
\ead{grzcen@amu.edu.pl}
\affiliation[inst1]{organization={Institute of Spintronics and Quantum Information, Faculty of Physics, Adam Mickiewicz University, Pozna{\'n}},
            addressline={Uniwersytetu~Pozna{\'n}skiego~2}, 
            city={Pozna{\'n}},
            postcode={61-614}, 
            country={Poland}}
            
\author[inst1]{Jaros{\l}aw W. K{\l}os}

\begin{abstract}

A planar square dot is one of the simplest structures confined to three dimensions. Despite its geometrical simplicity, the description of the spin wave modes in this structure is not trivial due to the competition of dipolar and exchange interactions. An additional factor that makes this description challenging are the boundary conditions depend both on non-local dipolar interactions and local surface parameters such as surface anisotropy. In the presented work, we showed how the surface anisotropy applied at the lateral faces of the dot can tune the frequency of fundamental mode in the planar CoFeB dot, magnetized in an out-of-plane direction. Moreover, we analyzed the spin wave profile of the fundamental mode and the corresponding dynamic stray field. We showed that the asymmetric application of surface anisotropy produces an asymmetric profile of dynamic stray field for square dot and can be used to tailor inter-dot coupling. The calculations were performed with the use of the finite-element method.

\end{abstract}



\begin{keyword}
square dot \sep boundary conditions \sep spin waves \sep finite-element method
\end{keyword}

\end{frontmatter}


\section{Introduction}
\label{sec:intro}
Flat dots and stripes are magnonic structures with non-trivial geometry, which can be relatively easily produced by patterning the magnetic layer \cite{Barman_2020,Rychly_2022,Gubiotti_2012,di_bona_magnetic_2007,Wang2010}. The magnetization dynamics in planar nanoelements (in flat dots and stripes) is clearly different from that observed in a infinite continuous film \cite{Pirro2021}. The obvious difference is the restriction of the propagation of the spin wave to one direction (in stripes) or its full confinement (in dots). The introduction of these geometrical constraints leads to the quantization of the wave vector in one or two dimensions, which qualitatively modifies the spectrum of spin waves \cite{Kra14_krawczyk2014review,Demokritov_magnonics_2013}. In the case of stripes, we are dealing with a spectrum in the form of a quasi-one-dimensional dispersion relation, where the propagation along the stripes is mediated by a series of mods resulting from the quantization of the wave vector across the stripes \cite{bessonov2015magnonic}. In dots, on the other hand, the dynamical magnetization is completely confined and the spectrum has a form of the set of discrete frequencies \cite{GUBBIOTTI_2007_sd}. In addition to the properties described above, which characterize any wave excitations, spin waves in nanostructures have specific properties. Their dynamics is limited only to the magnetic medium, and the boundary conditions at the surfaces with a non-magnetic medium result not only from the magnetic properties of the surface but are interdependent with the dynamics of magnetization inside the nanoelement \cite{Guslienko_2005}.
The parameter describing the freedom of magnetization precession on the surface of the magnetic body is the surface anisotropy \cite{Puszkarski1979} expressed by the surface anisotropy constant $K_{\rm s}$. 
This local parameter affects boundary conditions for dynamic component  $\mathbf{m}(\mathbf{r},t)=\mathbf{m}(\mathbf{r})e^{i\omega t}$ of magnetization $\mathbf{M}(\textbf{r},t)~=~\mathbf{M}_0(\mathbf{r})~+~\mathbf{m}(\mathbf{r},t)$ which take a general form (for exchange dominated spin waves)\cite{Gurevich1996}:
\begin{equation}\label{eq:BC}
\hat{\mathbf{z}}\times\frac{\partial \mathbf{m}}{\partial \hat{\mathbf{n}}}+\frac{K_{\rm s}}{A}\left(\left(\hat{\mathbf{n}}\cdot\mathbf{m}\right)\hat{\mathbf{n}}\times\hat{\mathbf{z}}+\left(\hat{\mathbf{n}}\cdot\hat{\mathbf{z}}\right)\hat{\mathbf{n}}\times\mathbf{m}\right)=0
\end{equation}
and can be used to modify the amplitude and ellipticity of spin wave precession at the surface.
For saturated sample in out-of-plane direction $\mathbf{M}_0=M_{\rm S}\hat{\mathbf{z}}$, where $M_{\rm S}$ saturation magnetization. The unit vector $\hat{\mathbf{n}}$ is normal to the surface and determines the axis of uniaxial anisotropy. The symbol $A$ denotes the exchange stiffness constant.
The sources of magnetic anisotropy (including surface  anisotropy) are attributed to spin-orbit interactions and are related to the deformation of electron shells by the crystal field in the bulk (and at the surface/interface). The anisotropy can be tuned by applying an electric field to the sample or by combining the ferromagnetic material with another material (e.g., heavy metal). Exemplary values of surface anisotropy constant can be found in \cite{GRADMANN_1986,Maruyama_2009}.

Formula (\ref{eq:BC}) does not take into account the influence of dipolar interactions, which, being long-range, are dependent on the size and shape of the nanostructure. Dynamic dipolar interactions result from both the inhomogenity of precession amplitude (volume charges) and the presence of surfaces (surface charges). In the case of planar structures, a reduction in the amplitude of the spin wave (so-called dipolar pinning) is observed at the side faces of the structure, depending on the width-to-thickness ratio. The dipolar pinning thus translates into boundary conditions for the spin waves \cite{Guslienko_2005}.

A peculiarity of planar magnonic structures is that the spin wave spectrum can be shaped by controlling the dipolar pinning at the surfaces of the structure through the selection of structural parameters \cite{centala2019influence, Guslienko_2005}, in addition to modifying the surface parameters (i.e., the surface anisotropy ). It is worth noting that in a system of nanoelements (e.g., the planar arrangement of strip or dot systems) separated by a non-magnetic material, a demagnetizing field ensures the coupling between them \cite{centala2019influence}. In magnetically saturated nanostructures, the dynamic demagnetizing field produced by fundamental mode (where the magnetization precesses in phase in the volume of the sample) significantly depends on the amplitude of magnetization near the surface, and this amplitude is usually reduced due to dipolar pinning, which is an obstacle to the design of dipolarly-coupled magnonic systems and devices of planar geometry \cite{Rana2013,Tacci2010,Keatley2008}.

In our work, we show that dipolar pinning and the associated effects: (i) an increase in the frequency of the fundamental mode and (ii) a reduction in the dynamic demagnetizing field, can be partially compensated by introducing uniaxial surface anisotropy on the lateral faces of the structure. We performed numerical studies for a planar square dot with a strong field applied perpendicularly to its plane. Our goal was also to show that by modifying the surface anisotropy on only one pair of lateral faces, it is possible to selectively reduce dipolar pinning and shape the dynamic demagnetizing field produced by the square dot independently in two perpendicular directions. This will allow the dot to dynamically couple to neighboring nanoelements with different strengths in two geometrically equivalent directions.

The layout of the work is as follows. In the subsequent section, we describe the considered structure and present the computational technique we used. Next, we discuss the results. We show that strong surface anisotropy on the lateral faces of the dot, can increase the precession amplitude at the edge of the dot and significantly reduce the frequency of the fundamental mode. We finish the results with a discussion concerning the spatial distribution and ellipticity of the dynamic demagnetization field produced by the fundamental mode. The paper concludes with a summary.


\section{Structure and methods}
\label{sec:MatMed}
\begin{figure}[!hb]
\includegraphics[width=0.95\linewidth]{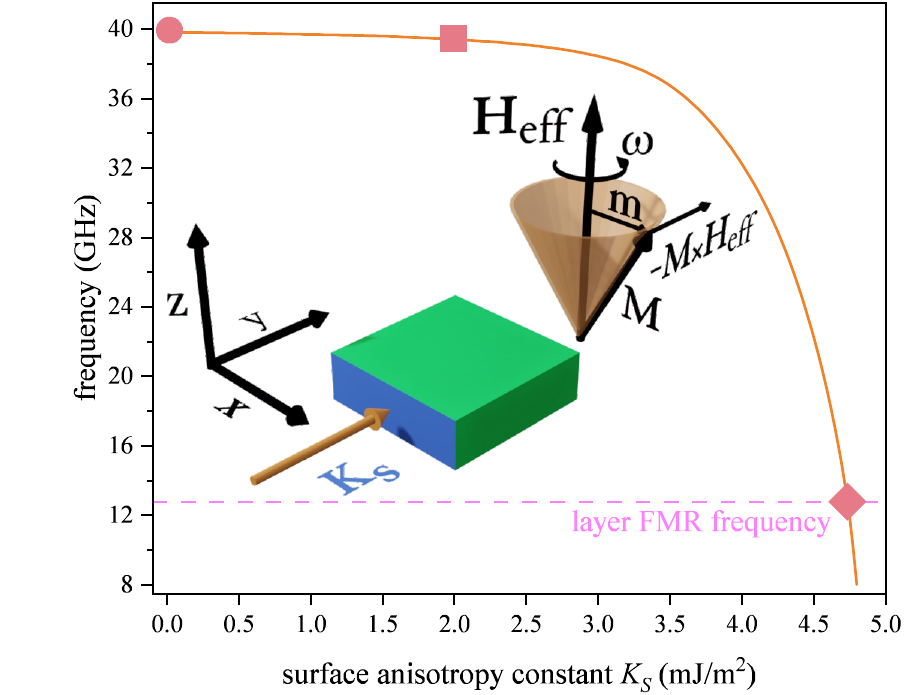}
\caption{
The frequency dependence on the surface anisotropy constant $K_{\rm s}$. The horizontal dashed line (magenta) indicates the frequency for the CoFeB layer in the forward volume configuration. Circle, square, and rhomb mark the points where the surface anisotropy ($K_{\rm s}$) takes values of 0, 2, and 4.735~mJ/m$^{2}$, respectively, considered further in the manuscript. 
An inset shows the square magnetic dot made of CoFeB with surface anisotropy ($K_{\rm s}$) on lateral sides for which the $y$-direction is normal. On the right of the considered structure, the cone of the magnetization precession
$\textbf{M}$ around the equilibrium position, given by static effective field  $\textbf{H}_{\rm eff}$. Magnetization spins at a circular frequency ($\omega$) and is forced by a torque equal $-\textbf{M}\times\textbf{H}_{eff}$.
}
\label{Fig:disprel_struct}
\end{figure}

The considered structure is a flat magnetic dot of square shape, made of cobalt iron boron alloy (CoFeB) with uniaxial surface anisotropy applied to the lateral surfaces for which $y$-direction is the normal direction (see inset Fig.~\ref{Fig:disprel_struct}). We consider the structure with the dimensions of 80~nm x 80~nm x 30~nm magnetized in a forward volume configuration. The material parameters used to perform the simulations were: magnetization saturation $\mu_0 M_{s}~=~1571$~mT, exchange stiffness constant $A~=~15$~pJ/m~\cite{Conca_2013}, and gyromagnetic ratio $\gamma~=~187$~rad/T/ns. The external magnetic field used for the simulation $\mu_0 H_{0}~=~2000$~mT is applied in the direction $\hat{\mathbf{z}}$, normal to the plane of the flat dot.

We obtained the results numerically using the finite element method. For this purpose, we used a semi-classical approach where the Landau-Lifshitz equation (\ref{eq:LLE}) describes the dynamics of the magnetization vector $\textbf{M}(\textbf{r},t)$ \cite{Gurevich1996}.
\begin{equation}\label{eq:LLE}
\frac{d\textbf{M}}{dt}=-\gamma\mu_{0}[ \textbf{M} \times \textbf{H}_{\rm eff} + \frac{\alpha}{M_{S}} \textbf{M} \times (\textbf{M} \times \textbf{H}_{\rm eff})]:
\end{equation}
where the symbol $\mu_{0}$ denotes the vaccum permeability,  $\alpha$ represents the damping constant, and $\textbf{H}_{\rm eff}(\textbf{r},t)$ is the effective magnetic field. In our studies, we neglect the damping constant since it is small $\alpha=0.0042$ ~\cite{Conca_2013}. 

We wanted to investigate the possibility of tuning the frequency of the dot fundamental mode by adjusting the surface anisotropy on its lateral faces. We also aimed to investigate the extreme case in which the fundamental modulus frequency is reduced to the frequency of ferromagnetic resonance in a uniform plane, even if the required values of the surface anisotropy constant are shifted beyond the experimentally verified range.

The effective magnetic field $\textbf{H}_{\rm eff}(\textbf{r},t)$ contains the following components: the external field  $H_{0}\hat{\mathbf{z}}$, exchange field $\textbf{H}_{\rm ex}(\textbf{r},t)= \left( 2A/\mu_{0} M^2_{s} \right)\laplacian\textbf{M}(\textbf{r},t)$, 
and dipolar field $\textbf{H}_{\rm d}(\textbf{r},t)=-\nabla 
\varphi (\mathbf{r},t)$~\cite{szulc2022}. The magnetostatic potential $\varphi$ is found using the Gauss law for magnetism and magnetostatic approximation \cite{Rychly_2022}.
We model the surface anisotropy by boundary conditions considering only the first component describing the uniaxial surface anisotropy, being in agreement with Eq.~\ref{eq:BC}.
 


\section{Results}
\label{sec:Results}

For the considered system, we decided to use a surface anisotropy constant exceeding typical experimental values ($K_{s}>1$~mJ/m$^{2}$ \cite{coey_2010}). We took this step to test whether the fundamental spin wave mode in a flat dot can be reduced to the FMR frequency of the unconstrained layer made of the same material. Using the Comsol Multiphysics, we have plotted the dependence of spin wave frequency on the anisotropy constant  (Fig.~\ref{Fig:disprel_struct}). The data indicate that such a reduction of frequency is theoretically possible. However, the required value of $K_{s}$  significantly exceeds the values known from experimental studies.

\begin{figure}[!ht]
\includegraphics[width=0.95\linewidth]{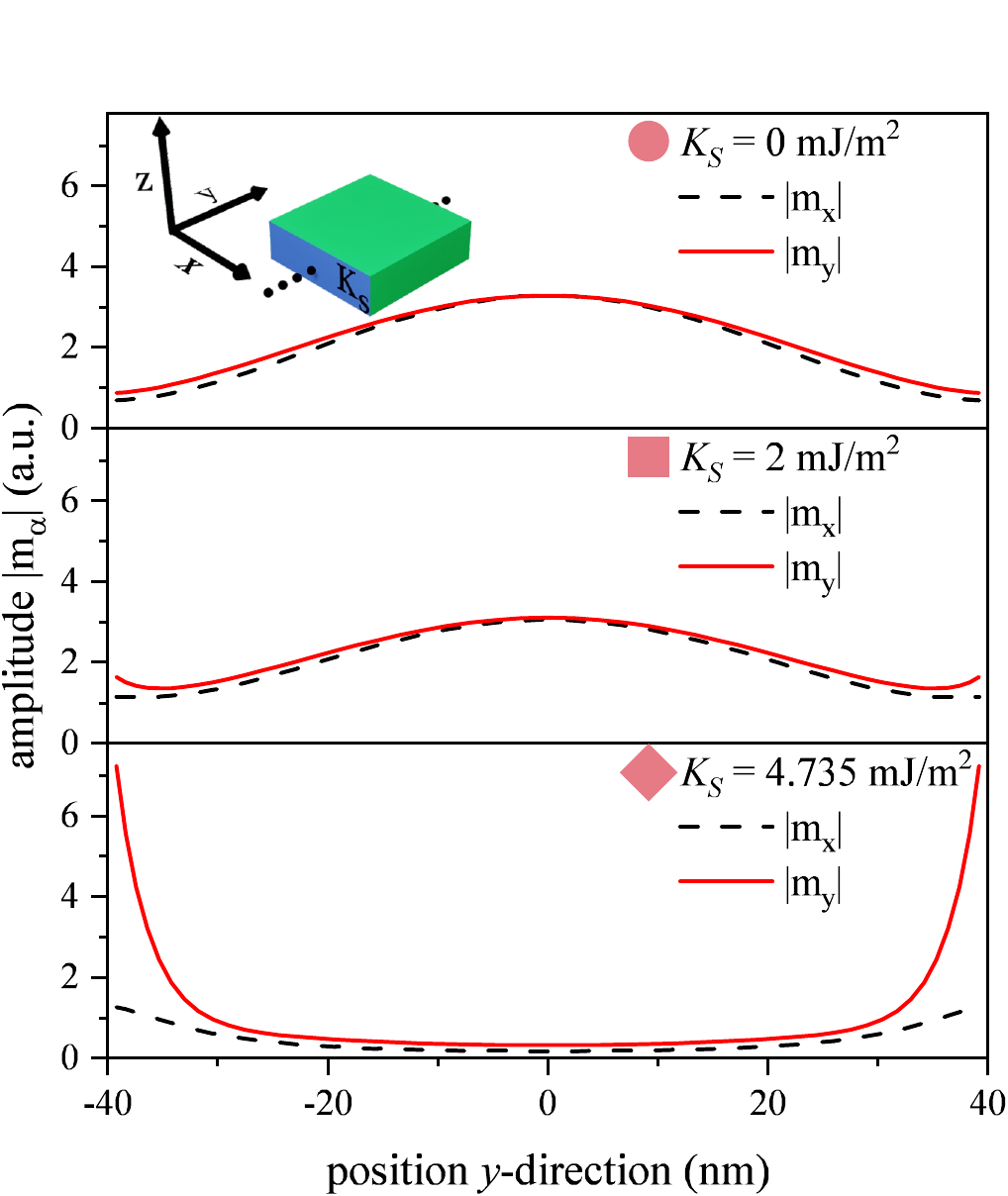}
\caption{
Amplitude profiles of the dynamic components of magnetization ($|m_{\alpha}|$, where $\alpha=x,y$) obtained by cutting the structure along the dotted line (see inset). Surface anisotropy was applied to the side walls of the structure for which the $y$-direction is normal. Profiles were obtained for surface anisotropy constant $K_{\rm s}= 0, 2, 4.735~{\rm mJ/m}^{2}$ -- see top, middle, and bottom plots, respectively. 
}
\label{Fig:mprofiles_struct}
\end{figure}

The relationship shown in Fig.~\ref{Fig:disprel_struct} cannot be considered in separation from the profiles shown in Fig.~\ref{Fig:mprofiles_struct}. For exchange dominated system, the absence of surface anisotropy $K_{\rm s}=0$ makes the magnetization free at the surface and the boundary  (\ref{eq:BC}) reduced to the form $\partial \textbf{m}/\partial \hat{\mathbf{n}}=0$. However, due to the presence of dipolar pinning \cite{Guslienko_2005}, the reduction of magnetization dynamics is observed at the lateral faces of flat dot and the spin waves profile is not uniform anymore -- see the top profile on Fig.~\ref{Fig:mprofiles_struct}.

By increasing the surface anisotropy constant, we release the normal component of dynamic magnetization at the edges to counteract dipolar pinning. The increase of the normal component of magnetization enhances the strength of the dynamic demagnetizing field, due to the increase of amplitude of surface charges \cite{centala2019influence}.
The tangential component of dynamic magnetization is also changed by the coupling resulting from the Landau-Lifshitz equation (see middle and bottom profiles on Fig.~\ref{Fig:mprofiles_struct}).

It is worth noting that the most rapid frequency drop in Fig.~\ref{Fig:disprel_struct} occurs for $K_{s}>3$~mJ/m$^{2}$. In this range,  the flatness of profile of the normal component of magnetization $m_y$ is not improving anymore. The limit of the FMR frequency of the film (dashed line in Fig.~\ref{Fig:disprel_struct}) is reached for highly non-uniform profile $m_y(y)$ with the amplitude of precession significantly uplifted at the edges of the dot -- see the bottom plot in Fig.~\ref{Fig:mprofiles_struct}).

\begin{figure}[!ht]
\includegraphics[width=0.95\linewidth]{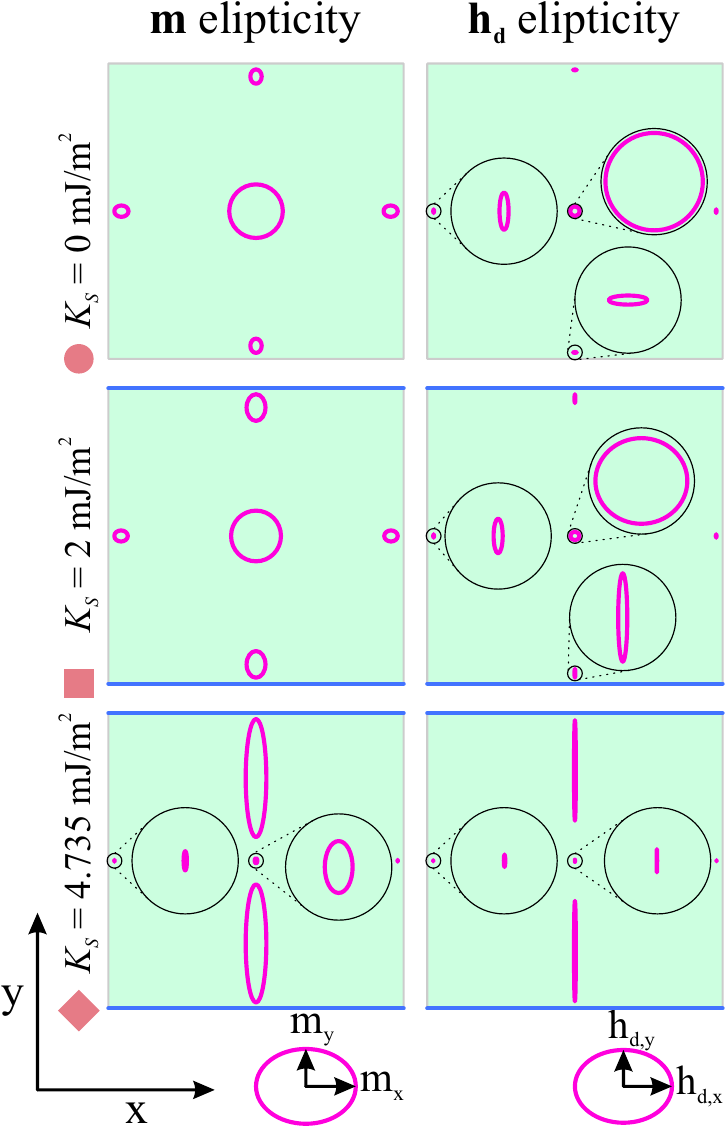}
\caption{
The ellipticity change associated with the modification of surface anisotropy $K_{\rm s}$ demonstrated for the dynamic components of magnetization $\textbf{m}$ and the dynamic magnetic field derived from dipolar interactions $\textbf{h}_{d}$. The results are presented for the values of surface anisotropy equal to 0, 2, and 4.735~mJ/m$^{2}$ for the upper, middle, and lower rows, respectively. The side walls of the structure on which the boundary conditions are affected by surface anisotropy $K_{\rm s}>0$ are marked with a bold blue line. The areas where the procession is weakly visible are magnified 10 times and presented in black circles.
}
\label{Fig:hmprofiles_struct}
\end{figure}

For $K_{\rm s}=0$, a circular ellipticity (see Fig.~\ref{Fig:hmprofiles_struct}) at the center of the square dot is evident for both the dynamic components of magnetization \textbf{m} and the demagnetization field \textbf{h}$_{\rm d}$. However, a discrepancy in ellipticity is apparent at the sides of the dot due to the demagnetizing effects related to the presence of the magnetic surface charges. The ellipse of the dynamic component of the demagnetization field is aligned with the long axis parallel to the nearest walls while the long axis of the magnetization ellipse is aligned perpendicular to the nearest walls.

For a surface anisotropy constant $K_{s}=2$~mJ/m$^{2}$, the ellipticity at the center of the dot for \textbf{m} and \textbf{h}$_{\rm d}$ remains almost circular while it changes near the edges of the dot. Near these edges where surface anisotropy has been applied, the long axis of the elliptical precession for \textbf{m} is elongated with respect to the $K_{s}=0$ state. At faces without surface anisotropy, this axis is shortened. When considering \textbf{h}$_{\rm d}$, please note that the direction of the long axis of ellipticity at the edges of the dot is oriented in parallel to the $y$-direction. 

For the case when $K_{\rm s}=4.735$~mJ/m$^{2}$ the long axes of both \textbf{m} and \textbf{h}$_{d}$ are directed parallel to the $y$ axis. Moreover, the ellipticity for \textbf{m} as well as \textbf{h}$_{d}$ at the edge of the dot where the surface anisotropy is applied becomes extremely large demonstrating the significant increase of the $m_{y}$ and $h_{{\rm d},y}$ components.

\begin{figure}[!ht]
\includegraphics[width=0.95\linewidth]{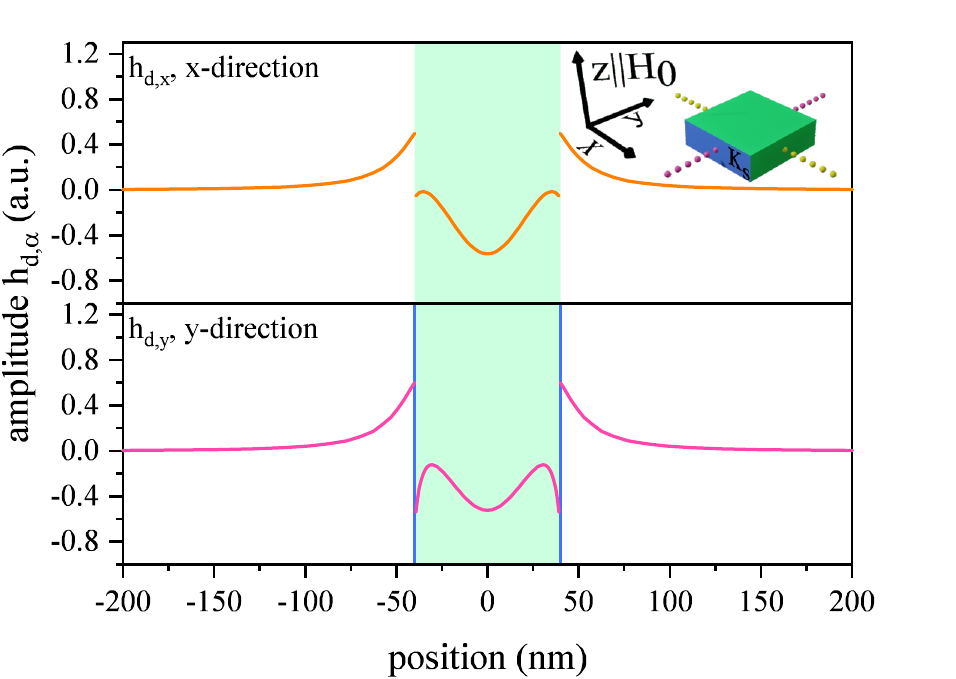}
\caption{
Profiles of the dynamic magnetic field component from dipolar interactions ($h_{{\rm d},\alpha}$, where $\alpha=x,y$) along $\alpha$-direction (see inset). The colors of dotted lines in the inset correspond to the color of $h_{{\rm d},\alpha}$ on the profiles. Profiles were made for the same value of surface anisotropy ($K_{\rm s}=2$~mJ/m$^{2}$) applied to the lateral surfaces perpendicular to $y$-direction. The lines on the green background correspond to the component of the dynamic magnetic field $h_{{\rm d},\alpha}$ inside the dot. The lines on the white background visualize the stray field. Bold blue lines represent surface anisotropy.
}
\label{Fig:hprofiles_struct}
\end{figure}


The dynamic components of magnetization and the dynamic components of the demagnetization field (inside the dot) increase the stray field (outside the dot) -- see Fig.~\ref{Fig:hprofiles_struct}. We can see the directional increase the stray field: $|h_{{\rm d},y}|>|h_{{\rm d},x}|$. As a result, such a dot, when placed in the square lattice, will interact differently in $x-$ and $y-$direction with its neighbors. In such a system, we should observe the difference in spin wave propagation in two principal directions $x$ and $y$, even if its geometry does not exhibit such asymmetry. 


\section{Conclusions}
\label{sec:Conclusions}
The magnetization in a confined structure is shaped by the interplay of dipolar and exchange interactions. However, for the structure of planar geometry, the surface anisotropy applied on lateral faces is an additional important factor which forms the profiles of confined modes, due to its impact on boundary conditions for spin waves. Moreover, it allows to shape the external dynamic magnetic field and thus the interaction with other dots of such kind.


\section{Acknowledgements}  
\label{sec:Acknowledgements}
This work has received funding from National Science Centre Poland grants  UMO-2020/39/O/ST5/02110,\break UMO-2021/43/I/ST3/00550 and support from the Polish National Agency for Academic Exchange grant BPN/PRE/2022/1/00014/U/00001.
\bibliographystyle{elsarticle-num} 
\bibliography{bibs.bib}





\end{document}